\documentclass[aps,prb,twocolumn,showpacs,amsmath,amssymb,superscriptaddress,reprint]{revtex4}
\usepackage{dcolumn}
\usepackage{bm}
\usepackage{graphicx}
\usepackage{epstopdf}
\usepackage{array}
\begin{document}

\title{Odd-frequency superconducting pairing in multi-band superconductors}
\author{Annica M. Black-Schaffer}
 \affiliation{Department of Physics and Astronomy, Uppsala University, Box 516, S-751 20 Uppsala, Sweden}
 \author{Alexander V. Balatsky}
 \affiliation{Nordic Institute for Theoretical Physics (NORDITA), Roslagstullsbacken 23, S-106 91 Stockholm, Sweden}
  \affiliation{Theoretical Division and Center for Integrated Nanotechnologies, Los Alamos National Laboratory, Los Alamos, New Mexico 87545, USA}

 \date{\today}

\begin{abstract}
We point out that essentially all multi-band superconductors have an odd-frequency pairing component, as follows from a general symmetry analysis of even- and odd-frequency pairing states. We show that odd-frequency superconducting pairing requires only a finite band hybridization, or scattering, and non-identical intraband order parameters, of which only one band needs to be superconducting. Under these conditions odd-frequency odd-interband pairing is always present. From a symmetry analysis we establish a complete reciprocity between parity in band-index and frequency.
\end{abstract}
\pacs{74.20.Rp, 74.45.+c, 74.78.Fk}
\maketitle

%
\section{Introduction}
One of the key aspects of superconductivity is the fermionic nature of the superconducting wave function, or equivalently the pair amplitude. This leads to the traditional classification into spin-singlet even-parity ($s$-, $d$-wave) or spin-triplet odd-parity ($p$-wave) pairing.

As Berezinskii \cite{Berezinskii74} originally showed, superconducting pairing can also be odd in time, or equivalently frequency. While theoretical proposals exist for odd-frequency bulk superconductors,\cite{Berezinskii74, Kirkpatrick91, Balatsky92, Abrahams95} odd-frequency pair amplitudes have so far only been argued to have been found in non-uniform systems, such as at surfaces and interfaces. \cite{Tanaka12}
For example, at superconductor-ferromagnetic interfaces a conventional spin-singlet $s$-wave superconducting pair amplitude is transformed into an odd-frequency spin-triplet $s$-wave amplitude, due to spin-rotational symmetry breaking.\cite{Bergeret01,Bergeret05} The spin-triplet nature gives rise to long-range proximity effect into the ferromagnet.
Also non-magnetic interfaces induce odd-frequency components, where instead translational symmetry breaking transforms a spin-singlet $s$-wave state into an odd-frequency spin-singlet $p$-wave state.\cite{Tanaka07JJ, Tanaka07PRB} The $p$-wave nature, however, makes this odd-frequency component sensitive to disorder.\cite{Tanaka07}

Numerous recently discovered superconductors have multiple bands at the Fermi level. These include both the unconventional  iron-pnictides/chalcogens,\cite{Kamihara08, Stewart11} heavy fermion superconductors,\cite{Stewart01, Rourke05, Seyfarth05} and MgB$_2$, a two-band phonon-driven superconductor.\cite{Choi02, Souma03} In these multi-band superconductors the band-index provides yet another symmetry index for the pair amplitude. While intraband pairing is, per definition, always an even function in band-index, both even- and odd-interband pairing are, in general, also possible.

In this work we show that odd-frequency pairing is ubiquitous in multi-band superconductors. By transforming between even- and odd-interband pairing, odd-frequency correlations are induced in the bulk of the superconductor, because the necessary symmetry breaking is, in general, present intrinsically in these systems. More specifically, we show that finite odd-frequency odd-interband pairing appears whenever there is a finite even-interband pairing between two non-identical bands. This is, for example, always the case when scattering, or hybridization, is present between two bands with non-identical intraband order parameters (of which one can be zero).
Formally, we find that the orbital, or band, parity (O) of the pair amplitude in multi-band superconductors, together with spatial parity (P) and time reversal (T), needs to obey the rule $PTO = +1 (-1)$ for spin-singlet (spin-triplet) pairing. There is thus a complete reciprocity between pairing that is odd in frequency and odd under band/orbital index permutation.

%
\section{Symmetry analysis}
We start by establishing the formal possibility of odd-frequency pairing in multi-band superconductors. The previous classification for even/odd-frequency pairing has to be broaden when multiple bands are present, as it is now also dependent on the orbital (band) parity. We generalize the Berezinskii approach \cite{Berezinskii74} by considering an orbital (or band or species) dependent two fermion condensate $\Delta_{\alpha \beta, a b}(r,\tau) = T_{\tau} \langle c_{\alpha a}(r, \tau) c_{\beta b}(0,0) \rangle$. Here $\alpha, a$ refer to spin and orbital index, respectively.
For concreteness we consider the case of two orbitals $a = 1,2$.
We define spatial parity (P) as acting on the relative coordinate $r$: $P \Delta_{\alpha \beta, a b}(r,\tau) = \Delta_{\alpha \beta, a b}(-r,\tau)$, time reversal (T) as acting on the relative time $\tau$: $T \Delta_{\alpha \beta, a b}(r,\tau) = \Delta_{\alpha \beta, a b}(r,-\tau)$, and {\em orbital parity} (O) as acting on the $a$ index: $ O \Delta_{\alpha \beta, a b}(r,\tau) = \Delta_{\alpha \beta, b a}(r,\tau)$.\footnote{For odd-orbital (antisymmetric) projection we trace $\Delta_{\alpha \beta}(r,\tau) = \sum_{ab} \varepsilon_{ab}\Delta_{\alpha \beta, a b}(r,\tau)$ and for even-orbital (symmetric) projection we trace $ \Delta^{I}_{\alpha, \beta}(r,\tau) = \sum_{ab}(\tau^I \cdot \tau^Y)_{ab}\Delta_{\alpha \beta, a b}(r,\tau)$, $I = 1,2,3$ is the vector index in the ``orbital" space" and $\tau$ are Pauli matrices operating in orbital space.}
The general symmetry requirement for a two fermion condensate can then be written as
\begin{equation}
\label{eq:def1}
PT \Delta_{\alpha \beta, a b}(r,\tau) = -\Delta_{\beta \alpha, b a}(r, \tau)
\end{equation}
For spin-singlet $S=0$ we further project to spin-singlet and find
\begin{equation}
 \Delta_{ab}(-r,-\tau)= \Delta_{ba}(r, \tau),
\end{equation}
which we shorthand as $PTO = 1$, namely, the simultaneous inversion of space, time, and permutation of orbital index will leave a spin-singlet pairing order parameter invariant.
For the spin-triplet case ($\Delta$ now is a vector in spin space) a similar analysis leads to
\begin{equation}
 {\vec{\Delta}}_{ab}(-r,-\tau)= -{\vec{\Delta}}_{ba}(r,\tau),
\end{equation}
which we shorthand as $PTO = -1$. The full symmetries of the two particle pair correlator are summarized in Table \ref{tab}.
\begin{table}[h]
 \setlength{\tabcolsep}{6pt}
 \begin{tabular}{| m{1.3cm} | c c c || m{1.3cm} |c c c|}
 \hline
    S = 0  &  P  &  T   & O
     &  S = 1   &   P   &  T & O \\
   \hline 

   even-$\omega$ & $+$ & $+$ & $+$ & even-$\omega$ & $-$ & $+$ & $+$ \\
   \hline
   even-$\omega$ & $-$ & $+$ & $-$ & even-$\omega$ & $+$ &$+$ & $-$ \\
   \hline
   odd-$\omega$ & $+$ & $-$ & $-$ & odd-$\omega$ & $+$ & $-$ & $+$ \\
   \hline
   odd-$\omega$ & $-$ & $-$ & $+$ & odd-$\omega$  & $-$ & $-$ & $-$ \\
   \hline
 \end{tabular}
 \caption{Behavior of the two fermion condensate under spatial parity (P), time-reversal (T), and orbital parity (O) symmetry for spin-singlet ($S = 0$, left), spin-triplet ($S = 1$, right) pairing, and different frequency ($\omega$) dependence.}
 \label{tab}
\end{table}
%
 
\section{Generic two-band superconductor}
The results in Table \ref{tab} provide the formal evidence of odd-frequency pairing in multi-band superconductors by changing the orbital (band) parity.
In order to show that odd-frequency pairing is also extremely common in multi-band superconductors we start with a generic two-band superconductor:
%
\begin{align}
\label{eq:Hab}
H_{ab}  & =  \sum_{{\bf k}\sigma} \varepsilon_a ({\bf k}) a^\dagger_{{\bf k}\sigma} a_{{\bf k}\sigma} + \varepsilon_b ({\bf k}) b^\dagger_{{\bf k}\sigma} b_{{\bf k}\sigma} \nonumber \\
& +  \sum_{{\bf k}\sigma} \Gamma({\bf k}) a^\dagger_{{\bf k}\sigma} b_{{\bf k}\sigma} + {\rm H.c.} \nonumber \\
& + \sum_{\bf k} \Delta_a({\bf k}) a^\dagger_{{\bf k}\uparrow} a^\dagger_{-{\bf k}\downarrow} + \Delta_b({\bf k}) b^\dagger_{{\bf k}\uparrow} b^\dagger_{-{\bf k}\downarrow} +{\rm H.c.}.
\end{align}
Here $a^\dagger_{{\bf k}\sigma}$ creates an electron in band $a$ with momentum ${\bf k}$ and spin $\sigma$, and similarly for band $b$. The kinetic energy is given by the band dispersions $\varepsilon_{a,b}$ and a single-particle band scattering, or hybridization, $\Gamma$. A finite $\Gamma$ appears automatically if the superconducting pairing occurs in (atomic or molecular) orbitals in which the kinetic energy is not fully diagonal,\cite{Japiassu92} as e.g.~proposed for the iron-pnictide superconductors.\cite{Moreo09b} It can also e.g.~result from disorder-induced interband scattering.
The superconducting intraband (diagonal) order parameters are $\Delta_{a,b}$. We will here assume conventional spin-singlet, uniform $s$-wave superconducting states, but the results apply equally well to any intraband pairing.
For finite $\Gamma$ we diagonalize the kinetic energy, resulting in a Hamiltonian with fully diagonal bands $c$ and $d$, but now with both intraband superconducting order parameters $\Delta_c$ and $\Delta_d$ and an even-interband order parameter $\Delta_{cd}$:
%
\begin{align}
\label{eq:Hcd}
H_{cd}  & =  \sum_{{\bf k}\sigma} \varepsilon_c ({\bf k}) c^\dagger_{{\bf k}\sigma} c_{{\bf k}\sigma} + \varepsilon_d ({\bf k}) d^\dagger_{{\bf k}\sigma} d_{{\bf k}\sigma} \nonumber \\
& + \sum_{\bf k} \Delta_c({\bf k})  c^\dagger_{{\bf k}\uparrow} c^\dagger_{-{\bf k}\downarrow} + \Delta_d({\bf k})  d^\dagger_{{\bf k}\uparrow} d^\dagger_{-{\bf k}\downarrow} +{\rm H.c.} \nonumber \\
& + \sum_{\bf k} \Delta_{cd}({\bf k})  (c^\dagger_{{\bf k}\uparrow} d^\dagger_{-{\bf k}\downarrow} + d^\dagger_{{\bf k}\uparrow} c^\dagger_{-{\bf k}\downarrow}) + {\rm H.c.}.
\end{align}
If we write $\Delta_b = \alpha \Delta_a$, we can express $\Delta_{cd} = (\alpha -1)\Delta_a |\Gamma|/\sqrt{(\varepsilon_a - \varepsilon_b)^2 + 4|\Gamma|^2}$. Thus, even-interband pairing is always present whenever $\Gamma \neq 0$ and $\Delta_a \neq \Delta_b$  in the original Hamiltonian $H_{ab}$. With the discovery of several multi-band superconductors, intrinsic even-interband pairing has also been studied in many systems.\cite{Dolgov87, TahirKheli98, Liu03, Gubankova03, Moreo09, Moreo09b,Ganguly11} In this case the need to start with a finite band hybridization $\Gamma$ in Eq.~(\ref{eq:Hab}) is automatically circumvent.

We are here primarily interested in the $s$-wave time-ordered pairing amplitude:
%
\begin{align}
\label{eq:Fpm}
F^{\pm}(\tau) = \frac{1}{2 N_{\bf k}}\sum_{\bf k} {\cal T}_{\tau}\langle c_{-{\bf k}\downarrow}(\tau) d_{{\bf k}\uparrow}(0) \pm  d_{-{\bf k}\downarrow}(\tau) c_{{\bf k}\uparrow} (0)\rangle,
\end{align}
which is an even ($+$) or odd ($-$) function in band index. $N_{\bf k}$ is the number of points in the first Brillouin zone.
$F^\pm(\tau)$ can also be either even or odd in the time coordinate. The even-frequency pair amplitude we define, as usual, by the equal-time amplitude, such that the even-frequency even-interband spin-singlet $s$-wave amplitude is $F^e = F^+(\tau \rightarrow 0^+)$.
The superconducting even-interband order parameter is thus $\Delta_{cd} = -U_{cd} F^e$, for some effective interband pairing potential $U_{cd}$. Since $\Delta_{cd}$ is an even function in band-index, the odd interband combination $F^o = F^-(\tau \rightarrow 0^+) = 0$.
For the component odd in time, we can still define an equal-time order parameter if we use the time derivative at equal times:\cite{Balatsky93, Abrahams95, Dahal09, Black-Schaffer12oddw} $F_\omega^o = \left. \frac{\partial F^-}{\partial \tau}\right|_{\tau \rightarrow 0^+}$. 
Odd-frequency pairing is necessarily accompanied by an oddness in band index for spin-singlet $s$-wave pairing.

To continue, we first focus on the interband pairing in $H_{cd}$ in Eq.~(\ref{eq:Hcd}), setting $\Delta_{c,d} = 0, \Delta_{cd} = \Delta$. Formally this can be achieved by choosing $\varepsilon_a = \varepsilon_b$ and $\Delta_a = -\Delta_b$.
By using the time-dependence $\gamma_i(\tau) = \gamma_i(0) e^{-iE_i\tau}$ of the $i$th Bogoliubov quasiparticle, with $E_i$ being its eigenenergy, we find
%
\begin{align}
\label{eq:Fwo}
F^o_\omega = \frac{i}{2N_{\bf k}} \sum_{\bf k} \frac{\Delta \left[\eta \sinh(\frac{\varepsilon_c-\varepsilon_d}{2{k_B T}}) + (\varepsilon_c-\varepsilon_d)\sinh(\frac{\eta}{2{k_BT}})\right]}
{\eta\left[ \cosh(\frac{\varepsilon_c-\varepsilon_d}{2{ k_BT}}) + \cosh(\frac{\eta}{2{k_BT}})\right]},
\end{align}
where $\eta = \sqrt{(\varepsilon_c + \varepsilon_d)^2 + 4|\Delta|^2}$. For odd-frequency pairing to appear $\varepsilon_c \neq \varepsilon_d$ is necessary, which is true for finite $\Gamma$. Further, when $T\rightarrow 0$ and $|\varepsilon_c-\varepsilon_d|>\eta$ we get $F_\omega^o = \frac{i}{2N_{\bf k}}\sum \frac{\Delta(\varepsilon_c-\varepsilon_d)}{\eta}$, whereas if $|\varepsilon_c-\varepsilon_d|<\eta$, $F_\omega^o = \frac{i}{2N_{\bf k}}\sum \Delta\,{\rm sgn}(\varepsilon_c-\varepsilon_d)$. Odd-frequency odd-interband pairing is thus always present in a superconductor with even-interband pairing and non-identical bands.\footnote{Technically both bands cannot also be half-filled, i.e.~particle-hole symmetry needs to be broken.} If there is no intrinsic even-interband pairing present, odd-frequency pairing will still always exist in a two-band superconductor with finite band hybridization $\Gamma$ and different intraband order parameters. The overall factor of $i\Delta$ in Eq.~(\ref{eq:Fwo}) is important as it gives $\pm [F^\pm(\tau)]^\ast = F(\tau)$ and thus invariance under time-reversal symmetry.

\begin{figure}[thb]
\includegraphics[scale = 1.05]{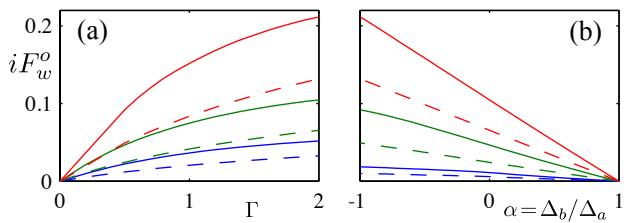}
\caption{\label{fig:bandhyb} (Color online) Odd-frequency odd-interband pair amplitude $F_\omega^o$ for a 3D two-band superconductor with $\varepsilon_a = -2\sum_{i} \cos(k_i a)$, $\varepsilon_b = \beta \varepsilon_a$ for $\beta = 1$ (solid), $\beta = 4$ (dashed) and $\Delta_a = 0.5$, $\Delta_b = \alpha \Delta_a$ in Eq.~(\ref{eq:Hab}). (a) $F_\omega^o$ as function of $\Gamma$ for $\alpha = 0.5$, 0, -1 (increasing values). (b) $F_\omega^o$ as function of $\alpha$ for $\Gamma = 0.1$, 0.5, 2 (increasing values).
}
\end{figure}

Equation~(\ref{eq:Fwo}) ignored intraband pairing. While these can change the value of the odd-frequency odd-interband pair amplitude they will, in general, never destroy it, as exemplified in Fig.~\ref{fig:bandhyb}.
There we plot $iF_\omega^o$ for a two-band superconductor on a three-dimensional (3D) cubic lattice with nearest neighbor hopping and $\varepsilon_b = \beta \varepsilon_a$ for $\beta = 1, 4$ and $\Delta_b = \alpha \Delta_a$ for $\Delta_a>0$, $|\alpha| \leq 1$. 
Let us first study the special case $\alpha = -1, \beta = 1$, which explicitly illustrates the deep connection between $F^e$ and  $F_\omega^o$. Then the diagonal band dispersions $\varepsilon_{c,d}  = \varepsilon_a \mp \Gamma$, intraband pairing $\Delta_{c,d} = 0$, and interband pairing $\Delta_{cd} = \Delta_a$. If we further assume $\Gamma <\Delta_{a}$, the even-interband pairing amplitude is the BCS gap equation: $F^e = -\frac{1} {2N_{\bf k}}\sum_{\bf k} \frac{\Delta_{a}}{\sqrt{\varepsilon_a^2 + |\Delta_a|^2}}$, whereas the odd-frequency odd-interband amplitude is $F_\omega^o = i\Gamma F^e$. 
Red solid curve in Fig.~\ref{fig:bandhyb}(a) shows the linear dependence on $\Gamma$ for $\Gamma <\Delta_a = 0.5$, while deviations from $\alpha = -1, \beta = 1$ give a sublinear dependence on the band hybridization $\Gamma$.
We also find a linear dependence on $\alpha$, as seen in Fig.~\ref{fig:bandhyb}(b), clearly demonstrating the robust dependence of $F_\omega^o$ on the interband pairing $\Delta_{cd} \propto (\alpha -1)$. The decrease in $F_\omega^o$ with increasing band-width ($\beta$) is also a sign of its connection to the even-frequency pair amplitude.

The above analytical and numerical results show that odd-frequency odd-interband pairing is extremely common in multi-band superconductors, requiring only a finite band hybridization and different intraband order parameters, as is generally always present. For example, in the presence of interband defect scattering, odd-frequency pairing should be present in the two-band superconductor MgB$_2$,\cite{Choi02, Souma03} high-temperature superconducting iron-pnictides/chalcogens,\cite{Stewart11} as well as in superconducting heavy fermion compounds.\cite{Stewart01, Rourke05, Seyfarth05}
The key to odd-frequency odd-interband pairing is the existence of even-interband pairing.
Interband pairing that is {\it not} an even function in band index will {\it not} have the same effect. For example, interband pairing of the form $c_{{\bf k} \uparrow}^\dagger c_{{\bf -k} \downarrow}^\dagger d_{{\bf -k'} \downarrow}  d_{{\bf k'} \uparrow}$, which constitutes an interband {\it pair} scattering mechanism,\cite{Suhl59} does not induce odd-frequency pairing.

The deep connection between parity in band index and frequency is further solidified if we consider the case of even-frequency odd-interband pairing, which for $s$-wave symmetry is necessarily a spin-triplet state. While such odd-interband pairing cannot be induced by simple band hybridization, it has been suggested for the iron-pnictides \cite{Dai08} and found in the proximity-induced superconducting response in topological insulators.\cite{Black-Schaffer13TISC}
Again, ignoring any intraband pairing $\Delta_{c,d}$ in $H_{cd}$ in Eq.~(\ref{eq:Hcd}) and replacing the even-interband pairing $\Delta_{cd}$ with an odd-interband spin-triplet term we arrive at an odd-frequency even-interband spin-triplet $s$-wave pairing amplitude exactly equal to the result in Eq.~(\ref{eq:Fwo}). Such mixing of even-frequency odd-interband and odd-frequency even-interband pairing has in fact been pointed out before for multi-pocket systems.\cite{Kubo08}

\section{Multi-orbital superconductors}

We have so far, exclusively worked in reciprocal space, but there are many situations where multiple superconducting orbitals, or sites, within one unit cell have to be described in real space. In this case we will let the operators $a_{i\sigma}$ and $b_{i\sigma}$ represent the (two) different orbitals in the unit cell $i$. By using $a_{i\downarrow}a_{i\uparrow}$ and $b_{i\downarrow}b_{i\uparrow}$ for intraorbital spin-singlet $s$-wave pairs, the derivation given above is equally applicable in this real space system.
Thus odd-frequency odd-interorbital pairing will always be present as soon as there is a finite single-electron orbital hybridization of the form $a_{i\sigma}^\dagger b_{i\sigma}$ and non-identical intraorbital superconducting order parameters. The latter requirement can be fulfilled if the orbitals have different physical origins, but also if the orbitals are separated in space and there are atomic-scale variations in the material. A Josephson junction with single-electron hybridization across the junction is a prototype example of the latter. This odd-frequency odd-interorbital pairing is very different from the odd-frequency pairing discussed earlier for Josephson junctions,\cite{Tanaka07JJ, Tanaka07PRB, Tanaka07} where a conventional spin-singlet $s$-wave junction generates odd-frequency spin-singlet $p$-wave pairing, which is not robust to disorder.
Another example is a superconductor/Bi$_2$Se$_3$ topological insulator heterostructure. The two active (Bi) orbitals in Bi$_2$Se$_3$ are separated along the $z$-axis \cite{Zhang09, Li10} and will therefore experience different superconducting pairing. We recently found numerically a complete reciprocity between parity in orbital and frequency spaces in a Bi$_2$Se$_3$/superconductor heterostructure for spin-singlet $s$-, $d$-wave as well as spin-triplet $p$-wave superconductors.\cite{Black-Schaffer13TISC} The $PTO = \pm 1$ symmetry requirement established above provide the analytical framework for this finding.

\begin{figure}[thb]
\includegraphics[scale = 1.03]{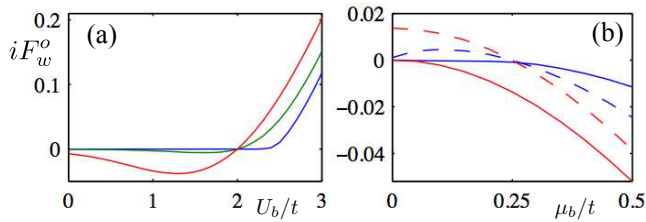}
\caption{\label{fig:graphene} (Color online) Odd-frequency odd-interband pair amplitude $F_\omega^o$ in graphene for nearest neighbor hopping $t = 2.5$~eV, chemical potentials $\mu_{a,b}$, and on-site pairing potentials $U_{a,b}$. (a) $F_\omega^o$ as function of $U_b$ for $U_a = 2t$ and $\mu_a = \mu_b = 0, 0.25t, 0.5t$ (increasing values). (b) $F_\omega^o$  as function of $\mu_b$ for $\mu_a = 0$ (solid), $0.25t$ (dashed) for $U_a= U_b = 2t, 3t$ (increasing values). Zero $F_\omega^o$ for $\mu = 0$ is caused by a superconducting quantum critical point at finite $U$.
}
\end{figure}

Yet another simple example of a multi-orbital system is graphene. Intrinsic superconductivity has been proposed theoretically in graphene \cite{Black-Schaffer07, Uchoa07, Gonzalez08, Nandkishore12} and a superconducting state has been achieved experimentally in graphene by proximity-coupling to a superconductor deposited on top of the graphene sheet.\cite{Heersche07}
In graphene, the hybridization between the $p_z$-orbitals on the two carbon atoms equals the nearest neighbor hopping $t$, and therefore overwhelmingly dominates kinetic energy.
Odd-frequency odd-interorbital pairing will thus be present whenever there are different superconducting intraorbital pairing order parameters $\Delta_{a,b}$ on the two sites.
Since $\Delta_a(i) = -U_a\langle a_{i\downarrow}a_{i\uparrow}\rangle$ for some pair potential $U_a$, and equivalently for $\Delta_b(i)$, different intraorbital order parameters can be achieved by either having $U_{a}\neq U_{b}$ or by having different density of states at each site. 
Such sublattice symmetry breaking effects can be present in both intrinsically superconducting graphene, due e.g.~to substrate effects, or at the graphene-superconductor interface when superconductivity is proximity-induced in the graphene.
In Fig.~\ref{fig:graphene} we plot $iF_\omega^o$ for both of these cases.
%
In Fig.~\ref{fig:graphene}(a) the pair potential $U_b$ is changed while $U_a$ and the local chemical potentials $\mu_a = \mu_b$ are kept fixed. The odd-frequency response is always zero for $U_a = U_b = 2t$ and is larger for higher chemical potentials, since larger density of states at the Fermi level cause larger even-interorbital pairing.
In Fig.~\ref{fig:graphene}(b) we instead set $U_a = U_b$ but vary the chemical potential difference between the two sites. $F_{\omega}^o$ is zero when there is no asymmetry between the two sites, i.e.~$\mu_a = \mu_b$, but is in general otherwise finite. 
The results in Fig.~\ref{fig:graphene} show that odd-frequency pairing is present as soon as there is sublattice symmetry breaking, which in graphene can e.g.~be achieved by substrate or interface effects.\cite{Giovannetti07}
While we have here used graphene as a simple example, odd-frequency odd-interorbital pairing will be present in any non-Bravais lattice with a site-dependent superconducting state. For these systems it is the sublattice symmetry breaking that facilitates the creation of odd-frequency pairing.

\section{Energy gap}
Odd-frequency superconducting pairing has in the past often been associated with the appearance of sub-gap states,\cite{Tanaka07JJ,Tanaka07PRB, Yokoyama07, Tanaka12, Asano13} or even a low-energy continuum.\cite{Balatsky92, Dahal09}
However, for odd-frequency odd-interband pairing, we do {\it not} in general find any low-energy states.
For the special case studied analytically above, i.e.~Eq.~(\ref{eq:Hcd}) with $\Delta_{c,d} = 0, \Delta_{cd} = \Delta_a$ and $\varepsilon_{c,d} = \varepsilon_a \mp \Gamma$, we find the eigen energies $E = \pm (\sqrt{\varepsilon_a^2 + \Delta_a^2} \pm \Gamma)$, and thus zero energy states for $\Gamma \geq \Delta_a$. However, there is no stable superconducting state for $\Gamma \geq \Delta_a$. The absence of pure interband superconductivity with zero energy states has also been established in other systems.\cite{Moreo09, Ganguly11}
The absence of sub-gap states is further confirmed by numerically solving the original Hamiltonian $H_{ab}$ in Eq.~(\ref{eq:Hab}). For isolated bands, i.e.~$\Gamma = 0$, we have the BCS energy gap relationship $E_g = \Delta_{a,b}^{sc}$ in each band, where the superscript $sc$ stands for the self-consistent result found for fixed pair potentials $U_{a,b}$. For finite $\Gamma$ we always find $E_g>{\rm min}(\Delta_a^{sc},\Delta_b^{sc})$, with $\Delta_{a,b}^{sc}$ modified in the presence of a finite band hybridization. Thus, the energy gap is never smaller than the intraband BCS gaps.
The lack of low-energy signatures of the odd-frequency odd-interband pairing is similar to the odd-frequency pairing behavior in topological insulator/superconductor heterostructures \cite{Black-Schaffer12oddwTI, Black-Schaffer13TISC} and in heavy-fermion compounds.\cite{Coleman93} Together these results demonstrate that odd-frequency pairing often have a frequency dependence which do not generate sub-gap states.

%
\section{Conclusions}
In summary, we have found the general symmetry rule for spatial parity $P$, time reversal $T$, and orbital parity $O$ for multi-band superconductors to be $PTO = 1 (-1)$ for spin-singlet (triplet) pairing.
Within a generic microscopic model of multi-band superconductors we have shown that odd-frequency pairing always exists in the form of odd-interband (orbital) pairing if there is any even-frequency even-interband pairing present, consistent with the general symmetry requirements. Even-interband pairing can exist intrinsically in multi-band superconductors, but also a finite band hybridization together with non-identical intraband order parameter strengths give even-interband pairing.
In fact, we find a complete reciprocity between parity in band (orbital) index and frequency for the superconducting pair amplitude, which naturally follows from $TO=1$ for spin-singlet $s$-wave (or spin-triplet $p$-wave) pairing. The $s$-wave nature makes the odd-frequency pairing resistant to disorder scattering. These results show that odd-frequency pairing is present in the bulk state of many superconductors, requiring no external symmetry breaking such as interfaces or magnetic fields.

In this work we assumed an even-frequency order parameter and showed that it induces an odd-frequency pair amplitude. An intriguing possibility is that the order parameter in some multi-band superconductors has an odd-frequency dependence, but that it induces a finite even-frequency pair amplitude, which is mistaken to also be the (even-frequency) order parameter. One example might be the heavy fermion compounds, which have been propose to have an odd-frequency order parameter.\cite{Coleman93} Since establishing the symmetry of the superconducting pair condensate is a crucial first step in elucidating the pairing mechanism, our work provides a  new approach to superconducting pairing in multi-band superconductors.

%
%
\begin{acknowledgments}
We are grateful to E.~Abrahams, D.~Scalapino and J.~R.~Schrieffer for numerous discussions on the nature of odd-frequency superconductivity and acknowledge funding from the Swedish research council (VR) and the European Research Council under the European Union's Seventh Framework Programme (FP/2007-2013)/ERC Grant Agreement DM-321031. Work at Los Alamos was supported by U.S.~DoE Basic Energy Sciences for the National Nuclear Security Administration of the U.S.~Department of Energy under contract DE-AC52-06NA25396.
\end{acknowledgments}


\end{document}